\documentstyle[12pt]{article}

\begin{document}
\input{epsf}
\setcounter{page}{1}

\pagestyle{plain}
\vspace{1cm}
\begin{center}
\Large{\bf A Possible Mechanism for Production of Primordial Black Holes}\\
\small \vspace{1cm}
{\bf Kourosh Nozari}\\
\vspace{0.5cm} {\it Department of Physics,
Faculty of Basic Science,\\
University of Mazandaran,\\
P. O. Box 47416-1467,
Babolsar, IRAN\\
e-mail: knozari@umz.ac.ir}
\end{center}
\vspace{1.5cm}
\begin{abstract}
Primordial Black Hole Remnants(PBHRs) can be considered as a primary
source of cold dark matter. Hybrid inflation provides a possible
framework for production of primordial black holes(PBHs) and these
PBHs evaporate subsequently to produce PBHRs.  In this paper we
provide another framework for production of these PBHs. Using
signature changing cosmological model and the generalized
uncertainty principle as our primary inputs, first we find a
geometric cosmological constant for early stage of universe
evolution. This geometric cosmological constant can lead to heavy
vacuum density which may be interpreted as a source of PBHs
production during the inflationary phase. In the next step, since it
is possible in general to have non-vanishing energy-momentum tensor
for signature changing hypersurface, this non-vanishing
energy-momentum tensor can be considered as a source of PBHs
production. These PBHs then evaporate via the Hawking process to
produce PBHRs. Finally, possible observational schemes for detecting
relics of these PBHRs are discussed.
\end{abstract}
\newpage
\section{Motivation}
Dark matter constitutes a great part of the matter in the universe.
There are several candidates for dark matter[1]. Recently, black
holes remnants have been considered as a possible source of cold
dark matter[2,3]. In fact, Generalized Uncertainty Principle(GUP)
prevents black holes from total evaporation and the ultimate phase
of an evaporating black hole is a Planck size remnants. Evaporating
Primordial black holes provide possible framework for such Planck
size remnants[4]. Possible creation of black hole remnants in
inflationary scenarios, ultrahigh energy cosmic ray (UHECR)
airshowers and also in Large Hadronic Collider(LHC) have been
investigated[5-7]. Here we are going to provide another model for
creation of evaporating primordial black holes based on quantum
cosmological considerations. The possible production of PBHRs in
Hartle-Hawking quantum cosmology first has been emphasized by Chen
and Adler[4]. However, they have not provided a thorough formulation
of their conjecture. Here we are going to provide a framework for
possible production of PBHs in Hartle-Hawking scenario of quantum
cosmology. Our calculation is based on ref.[8]. First we will
construct a signature changing cosmological model. Then a
geometrical interpretation of the cosmological constant will be
given based on the generalized uncertainty principle. This
cosmological constant can be considered as a heavy vacuum density
which may be possible source of PBHs production. Finally we will
show that non-vanishing energy-momentum tensor of change surface may
also be a possible source of PBHs production. The essence of our
analysis is the fact that quantum fluctuations of spacetime geometry
in Planck scale may produce non-vanishing cosmological constant and
non-vanishing energy-momentum tensor(due to non-vanishing jump of
the extrinsic curvature) and these entities can be considered as
possible sources of primordial black hole production. We use the
thin shell formalism of general relativity within Colombeau new
theory of generalized functions. We also use the generalized
uncertainty principle(GUP), existence of minimal length scale and
the Hartle-Hawking no-boundary proposal as our primary inputs.\\
The structure of the paper is as follows: section 2 is devoted to
provide our model based on signature change and the generalized
uncertainty principle. In this section we will find a cosmological
constant with geometric origin which can lead to heavy vacuum
density deriving inflation. The possible non-vanishing of
energy-momentum tensor of signature change surface due to
non-vanishing jump of the extrinsic curvature is described as a
source of PBHs production. In section 3, following the arguments of
Chen and Adler[4], we investigate the possible production of PBHRs
in the signature change framework. Experimental and observational
schemes for detection of PBHRs are discussed in section 4. The paper
follows by summary and conclusions in section 5.\\

\section{Signature Change, GUP and the Cosmological Constant}
The idea of signature change has been originated from the
Hartle-Hawking No-Boundary proposal[9]. The essence of the idea of
signature change can be summarized as follows:\\
The universe was initially Riemannian $4$-sphere with Euclidean
signature $(+,\,+,\,+,\,+)$ of metric, and then by a change of
signature of "spacetime" metric(via a quantum tunneling process),
the transition to usual Lorentzian(de Sitter) spacetime with metric
signature of $(-,\,+,\,+,\,+)$ have been occurred. The transition
had been occurred on a hypersurface with $3$-sphere topology.\\
Earlier attempts to describe this proposal was based on Euclidean
path integral formulation of quantum gravity and the analogy to
quantum tunneling effect in quantum mechanics[10], but it has been
revealed that the idea of signature change can be formulated in the
language of classical general relativity[11]. Among these
literature, Mansouri and Nozari have constructed a cosmological
model with signature change in the framework of Colombeau
algebra[8,12]. In which follows, we use Mansouri-Nozari results to
provide our model of PBHs production.
\subsection{A Geometric Cosmological Constant}
According to the Hartle-Hawking no-boundary proposal, the universe
after signature change should be a de Sitter universe (inflationary
phase). Let us assume that the space time after signature change is
a de Sitter one. Consider now the following de Sitter metric with
appropriate lapse function $f(t)$ in order to produce signature
change at $t=0$. The $t = const.$ sections of this metric are
surfaces of constant curvature $k = 1$ [13]
\begin{equation}
\label{math:5.46} ds^2 = -f(t)dt^2 + a^2(t)\Bigg( d\chi^2 +
\sin^{2}\chi(d\theta^2 + \sin^2\theta d\theta^2)\Bigg)
\end{equation}
where $f(t)$ and $a(t)$ are defined as
\begin{equation}
\label{math:2.1} f(t) =  \theta (t) - \theta (-t)
\end{equation}
and
\begin{equation}
\label{math:5.47} a^2(t) =
\alpha_{+}^{2}\cosh^{2}(\alpha_{+}^{-1}t)\theta(t) +
\alpha_{-}^{2}\cos^{2}(\alpha_{-}^{-1}t)\theta(-t).
\end{equation}
Suppose that $[a] = 0$, therefore we will have $\alpha_{+} =
\alpha_{-} :=R$, where $R$ is the radius of 3-surface of signature
change . Now, the Euclidean sector can be interpreted as a $S^4$
with $S^3$ sections defined by $t = const$. The boundary of the
Euclidean sector, defined by $t = 0$, is a $S^3$ having the radius
$R = \alpha_{-} = H_{0}^{-1}$ which is the maximum value of
$\alpha_{-}\cos(\alpha_{-}^{-1}t)$ (see Fig. 1). In the Lorentzian
sector the cosmological constant is given by $\Lambda =
3\alpha_{+}^{-2} = 3 H_{0}^2$. The $t=const.$ surfaces are $S^{3}$
with radius $\alpha_{+}\cosh(\alpha_{+}^{-1}t)$ having the minimum
value $R = \alpha_{+} = H_{0}^{-1}$ (see Fig.1). Therefore, the
following relation between the cosmological constant and the radius
of the boundary is obtained:
\begin{equation}
\label{math:5.48} \Lambda = \frac{3}{R^2}.
\end{equation}
According to the string theory and loop quantum gravity, there is a
minimum length scale of the order of Planck length which restricts
the accuracy of measurement of distances[14]. In other words, one of
the most interesting consequences of unification of gravity and
quantum mechanics is that in resulting quantum gravity scenario,
there exists a minimal observable distance on the order of the
Planck length, $ l_{P} =\sqrt{\frac{G\hbar}{c^3}}\sim 10^{-33}cm$,
where G is the Newton constant. The existence of such a fundamental
length is a dynamical phenomenon due to the fact that, at Planck
scale, there are fluctuations of the background metric, i.e. a limit
of the order of Planck length appears when quantum fluctuations of
the gravitational field are taken into account. In the language of
string theory one can say that a string cannot probe distances
smaller than its length. The noncommutativity of spacetime at
quantum gravity level supports this idea[15]. This minimal length
scale can be obtained from the following generalized uncertainty
principle
\begin {equation}
\label{math:1 .1} \Delta x \Delta p \geq \frac{\hbar}{2}
(1+\beta(\Delta p)^2)
\end {equation}
which yields $(\Delta x)_{min}=\hbar\sqrt{\beta}$ and this is on the
order of Planck length [16]. Now it is reasonable to set $R\sim
l_{P}$ and therefore
\begin{equation}
\label{math:5.48} \Lambda \sim \frac{3}{l_{P}^2}.
\end{equation}
This statement shows the possibility of having an exponentially
expanding universe with heavy vacuum density probably equal to the
Planck density. This heavy vacuum density can be considered as a
source of PBHs production during inflationary era[17].

\subsection{Non-Vanishing Energy-Momentum Tensor of Signature Change
Hypersurface} Following the formalism of shell dynamics in general
relativity and using Colombeau algebra [18], one can show that
elements of energy-momentum tensor of the signature change
hypersurface are
\begin{equation}
\label{math:5.49} \kappa S_{\mu}^{\nu} = diag\Big( 0,\,\,
\Upsilon,\,\, \Upsilon,\,\, \Upsilon\Big),
\end{equation}
where $\Upsilon$ is defined as
\begin{equation}
\label{math:5.50} \Upsilon = \left(\frac{2}{R}\tanh(R^{-1}t) -
\frac{2}{R}\tan(R^{-1}t)\right)|_{(t = 0)}(\tau-1)
\end{equation}
where $\frac {1}{2}<\tau<1$ is a constant [8]. We therefore conclude
that given the de Sitter metric in the form of (1) the
energy-momentum tensor of the hypersurface of signature change
defined by $t = 0$ vanishes. However, this is not a general
statement since one could require a matching along other sections
corresponding to a non-maximum radius of the Euclidean sector or a
non-minimum radius of the Lorenzian sector. Within the
distributional approach, using Coloumbeau algebra, we obtain in
general a non-vanishing expression for the energy-momentum tensor
$S_{\mu\nu}$. In another words, one could require a matching along
other sections corresponding to a non-maximum radius of the
Euclidean sector (see Fig. 2) or a non-minimum radius of the
Lorenzian sector (see Fig. 3). In these situations, energy-momentum
of the change surface is non-vanishing. We consider this
non-vanishing energy-momentum tensor as a possible source of
primordial black holes production . Note that generalized
uncertainty principle itself is a consequence of quantum fluctuation
of background spacetime geometry . As has been indicated, in the
spirit of path integral approach, Hartle-Hawking proposal is on the
basis of quantum fluctuation of geometry itself. Therefore, one can
argue that quantum fluctuation of geometry (which are considered in
path integral) are the source of non-vanishing energy-momentum
tensor( non-vanishing jump of the extrinsic curvature in the
Darmois-Israel approach[19]) for change surface. This non-vanishing
energy-momentum tensor can be considered as a source of BPHs
production. Note that these two issues: geometric cosmological
constant and non-vanishing energy-momentum tensor are related via
the junction conditions which can be derived from full field
equations. In the next section we discuss the possible creation of
primordial black hole from these two features of our model:
geometric cosmological constant and non-vanishing energy-momentum
tensor of the signature change hypersurface .

\section{Signature Change and Primordial Black Hole Production}
As has been indicated, black hole remnants (BHRs) are a natural
candidate for cold dark matter since they are a form of weakly
interacting massive particles (WIMPs). The possible source and
abundance of BHRs are of interest. The most natural source is in
primordial geometric fluctuations, which can be sufficiently large
only in the Planck era, at about the Planck temperature. In previous
sections we have provided a consistent framework for such geometric
fluctuations in Planck era. Here, using analysis of Chen and
Adler([4]and [17]) we address the possible PBHs production in the
very early universe via the signature change. Rigorous
derivations(for example analysis provided by Ohanian and
Ruffini[20]) and also simple thermodynamic arguments( see for
example [21]) imply that random fluctuations can produce a Boltzmann
distribution of black holes, down to Planck mass, with a number
density of $\sim \frac{1}{l_{p}^{3}}$. As Ohanian and Ruffini have
shown[20], in one version of standard inflationary cosmology the
scale factor can increase by a factor of about $10^{74}$ from the
Planck era to the present, and since the number density of matter is
related to the cube of this, we obtain a number density about
$10^{-118}/m^3$. But the large scale density of dark matter is
approximately equal to the critical density, that is, $\rho_{DM}\sim
\rho_{c} \sim 2\times 10^{-26}kg/m^{3}$. This results a BHR number
density $\sim 10^{-18}/m^{3}$ which is impossible. In last sections
we have provided a framework for an alternative signature changing
cosmology which can support primordial black holes production. Using
Hartle and Hawking no-boundary scenario for creation of the
Universe, we have supposed that the universe was initially a truly
chaotic quantum foam system without ordinary spacetime, and in
particular without a time direction such that the signature was
$(+,\,+,\,+,\,+)$. A fluctuation in the signature to
$(-,\,+,\,+,\,+)$ can then produce a time direction and turn it into
an exponentially expanding de Sitter space with heavy vacuum
density(geometric cosmological constant) which can be equal to the
Planck density. In this framework, at the beginning of the universe
a thermal distribution of black holes can be produced as has been
pointed by Ohanian and Ruffini[20]. During the expansion, these
thermal black holes can decay to form Planck size remnants and
radiation. The presence of the black holes and radiation (photons,
gravitons, etc.) changes the equation of state. Heavy
vacuum(geometric cosmological constant) equation of state with $p
=-\rho$, changes to a mixture of BHRs matter with $p = 0$ and
radiation with $p = \rho/3$( it is possible to have a very little
residual vacuum energy also). This change of equation of state can
change the scale factor from an exponential to a power-law form. In
this situation apparently there is no horizon paradox[4,17].

Suppose that this transition occurs at a continuous energy density
change. As a matching condition for creation of de Sitter phase, the
scale factor should be continuous with continuous derivatives.
Therefore, the scale factor can change according to following
transformation
\begin{equation}
\label{math:3.1} a(t)=e^\frac{t}{t_{p}} \quad \Rightarrow \quad
a(t)=(\frac{e}{n})^{n} \frac{t^n}{t_{p}^{n}}
\end{equation}
at $t = nt_{p}$, where $n$ should be between 1/2 for radiation and
2/3 for matter. The duration of exponential expansion will be quite
short. One can estimate approximately the decrease in number density
of BHRs from the beginning of time, $t = 0$, to the present time, $t
= t_0$, as follows
\begin{equation}
\label{math:3.1}\Big[\frac{a(t_{0})}{a(0)}\Big]^{3}\approx
\Big[\frac{t_{0}}{t_{p}}\Big]^{3n}
\end{equation}

Suppose that $n$ is about $2/3$(which is appropriate for matter),
then the scale factor decreases by a factor of about $10^{41}$ and
correspondingly,  the density factor can decrease by a factor of
about $10^{123}$. This implies a present number density of black
hole remnants of order $\rho_{BHR}\approx 10^{-18}m^{-3}$. This
amount is reasonable. But the universe evolution contains a
radiation dominated era also and we should consider it in our
estimations. If we take into account a radiation dominated period of
expansion, then $n$ should be about $1/2$ until the time of
decoupling of matter and radiation, $t_d$, and the present density
should be about $10^{10}/m^3$, which is very large to be reasonable.
To overcome this difficulty, as has been pointed in [4,17], we have
to consider an ad hoc period of inflation to obtain a reasonable
density. By taking this strategy, if we extend the period of
inflation from $n t_p$ to $\eta t_p$, followed by a period of
radiation dominance to $t_d$, and then matter dominance to the
present, we obtain[4]
\begin{equation}
\label{math:3.1}\frac{a(t_{0})}{a(0)}\approx
\frac{e^{\eta}}{\sqrt{\eta}}\Big(\frac{t_{0}^{2/3}}{t_{p}^{1/2}t_{d}^{1/6}}\Big)
\end{equation}
This gives a suitable PBH density if we chose the number of
$e$-folding to be about $\sim 27$, roughly half the number usually
used in standard inflationary scenarios. The quantum field
theoretical framework for actual mathematical formalism can be
performed which is out of purpose of the present work. In which
follows we discuss the important issue of possible detection of
primordial black hole remnants.

\section{Schemes for Possible Detection of Primordial Black Holes Remnants}
It is by now widely accepted that dark matter (DM) constitutes a
substantial fraction of the present critical energy density in the
Universe. However, the nature of DM remains an open problem. There
exist many DM candidates, most of them are non-baryonic weakly
interacting massive particles (WIMPs), or WIMP-like particles. By
far the DM candidates that have been more intensively studied are
the lightest supersymmetric (SUSY) particles such as neutralinos or
gravitinos, and the axions (as well as the axinos). There are
additional particle physics inspired dark matter candidates. A
candidate which is not as closely related to particle physics is the
relics of primordial black holes(Micro Black Holes)which we have
investigated its production. Certain inflation models naturally
induce a large number of such a black holes. As a specific example,
hybrid inflation can in principle yield the necessary abundance of
primordial black hole remnants for them to be the primary source of
dark matter. One of the major problems with these remnants is the
possibility of their detection. As interactions with black hole
remnants are purely gravitational, the cross section is extremely
small, and direct observation of these remnants seems to be
unlikely. One possible indirect signature may be associated with the
cosmic gravitational wave background. Unlike photons, the gravitons
radiated during evaporation would be instantly frozen. Since,
according to our notion, the black hole evaporation would terminate
when it reduces to a remnants, the graviton spectrum should have a
cutoff at Planck mass. Such a cutoff would have by now been
red-shifted to $\sim10^{14} GeV$. Another possible gravitational
wave-related signature may be the gravitational wave released during
the gravitational collapse. The frequencies of such gravitational
waves would by now be in the range of $\sim 10^{7} - 10^{8} Hz$. It
would be interesting to investigate whether these signals are in
principle observable. Another possible signature may be some
imprints on the cosmic microwave background(CMB) fluctuations due to
the thermodynamics of black hole remnants-CMB interactions. Possible
production of such remnants in Large Hadronic Collider (LHC) and
also in ultrahigh energy cosmic ray (UHECR) air showers are under
investigation. If we consider hybrid inflation as our primary
cosmological model, there will be some observational constraints on
hybrid inflation parameters. For example a simple calculation based
on hybrid inflation suggests that the time it took for black holes
to reduce to remnants is about $10^{-10} Sec$. Thus primordial black
holes have been produced before baryogenesis and subsequent epochs
in the standard cosmology[17]. The events that can potentially lead
to black hole production are essentially high-energy scattering in
particle colliders and UHECR. The next generation of particle
colliders are expected to reach energies above $10$ $TeV$. For
example, LHC is planned to reach a center-of-mass energy of $14
TeV$. Therefore, if the fundamental Planck scale is of the order of
few $TeV$, LHC would copiously produce black holes. Black hole
production by cosmic rays has also been recently investigated by a
number of authors(see for example[22,23] and references therein).
Cosmogenic neutrinos[6] with energies above the Greisen-
Zatsepin-Kuzmin (GZK) cutoff[7] are expected to create black holes
in the terrestrial atmosphere. The thermal decay of the black hole
produces air showers which could be observed. The cross sections of
these events are two or more orders of magnitude larger than the
cross sections of standard model processes. Therefore, black holes
are created uniformly at all atmospheric depths with the most
promising signal given by quasi-horizontal showers which maximize
the likelihood of interaction. This allows black hole events to be
distinguished from other standard model events. Detecting $TeV$
black hole formation with UHECR detectors may be possible through
the decay of $\tau$-leptons generated by $\nu_\tau$'s that interact
in the Earth or in mountain ranges close to the detectors. A
secondary $\tau$ generated through the decay of a black hole has
much less energy than the standard model $\tau$ secondary. In
addition, black holes may produce multiple $\tau$-leptons in their
evaporation, a unique signature of $TeV$ gravity. Standard model
processes that generate multiple $\tau$-leptons are highly unlikely,
the detection of multiple $\tau$'s in earth-skimming and mountain
crossing neutrinos will be a smoking gun for black hole formation.\\

\section{Summary}
This paper provides a possible mechanism for primordial black hole
production in early universe. The arguments presented here are based
on the idea of signature change. It has been shown that junction
condition for signature change can leads to a geometrical
cosmological constant. When we consider the existence of a minimal
length scale on the order of Planck length, this geometric
cosmological constant can be interpreted as a heavy vacuum density
and this heavy vacuum density can be considered as a possible source
of PBHs production. From another viewpoint, possible non-vanishing
of energy-momentum tensor of the signature change hypersurface due
to matching along non-maximum (or non-minimum) radius, can leads a
possible source of PBHs production. These PBHs evaporates via
Hawking process to reach eventually to a Planck size black hole
remnants and these PBHRs can be interpreted as a source of dark
matter. Possible experimental and observational schemes for
detection of
these primordial black hole remnants are discussed also.\\

\begin{figure}[ht]
\begin{center}
\includegraphics{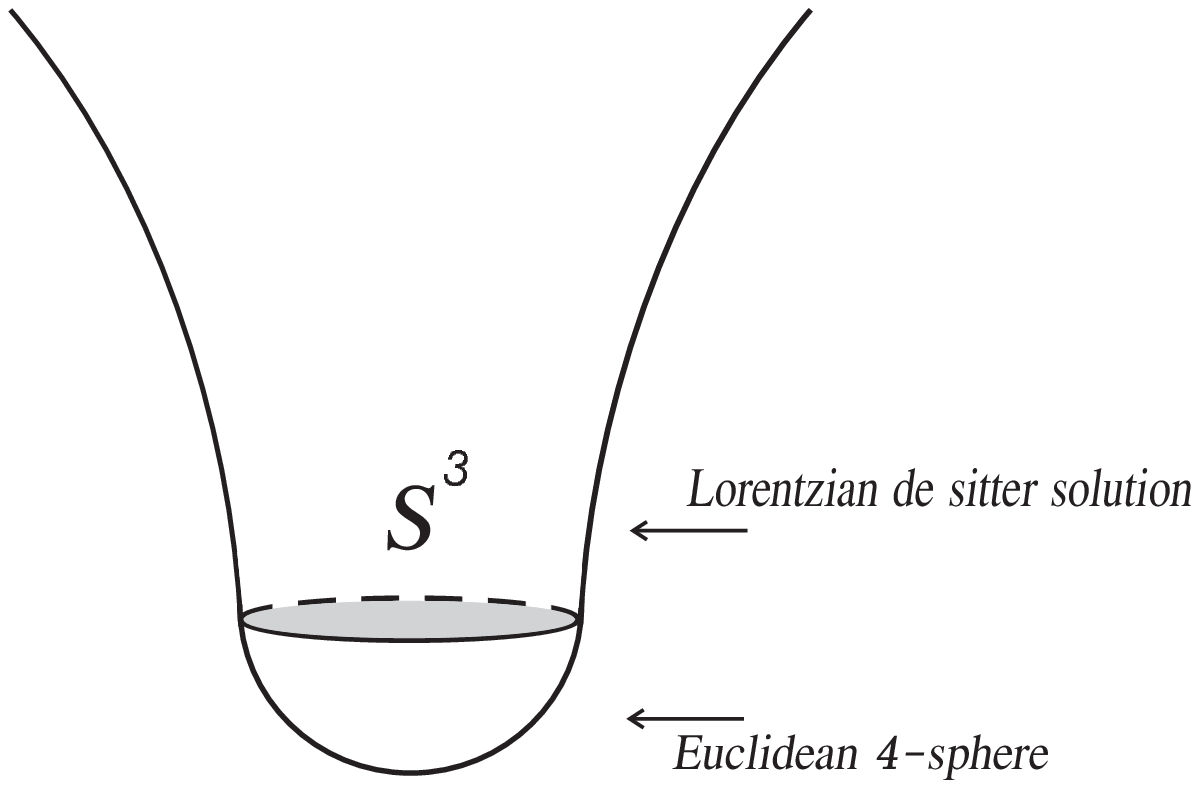}
\end{center}
\vspace{16 cm} \caption{\small {Creation of de Sitter Spacetime }}
\label{fig:1}
\end{figure}

\begin{figure}[ht]
\begin{center}
\includegraphics{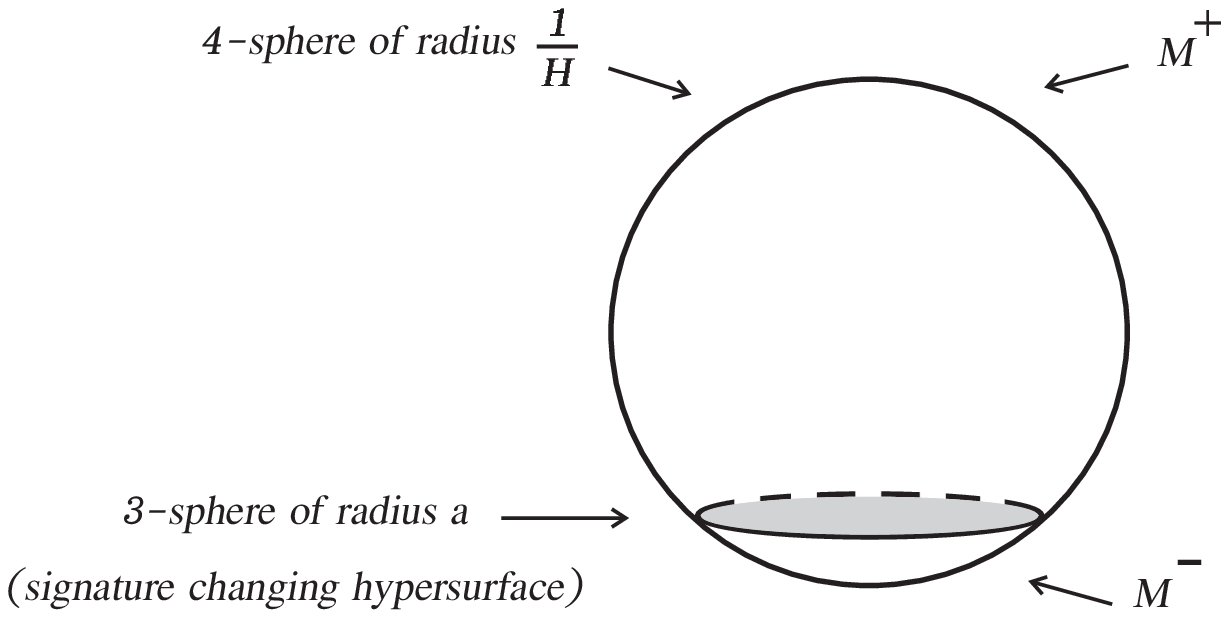}
\end{center}
\vspace{16 cm} \caption{\small {Matching on the signature changing
hypersurface for the case of non-maximum radius in Euclidean sector
}} \label{fig:2}
\end{figure}

\begin{figure}[ht]
\begin{center}
\includegraphics{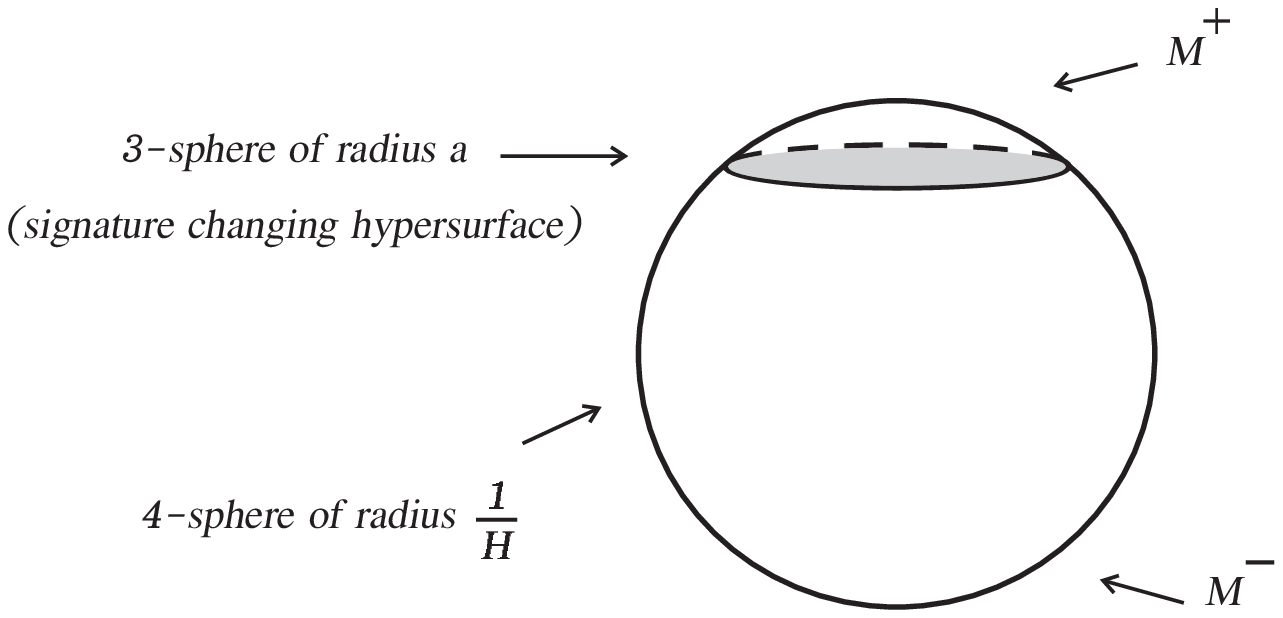}
\end{center}
\vspace{16 cm} \caption{\small {Matching on the Signature Changing
Hypersurface for the case of non-maximum radius in Lorentzian
sector. }} \label{fig:3}
\end{figure}

\end{document}